\documentclass[11pt]{article}
\usepackage[margin=1in]{geometry}
\usepackage[utf8]{inputenc}
\usepackage{amsmath}
\usepackage{amssymb}
\usepackage{amsfonts}
\usepackage{amsthm}
\usepackage{graphicx}
\usepackage{url}
\usepackage[super,numbers,sort&compress]{natbib}
\usepackage{graphicx}
\usepackage{dcolumn}
\usepackage{bm}
\usepackage{xr-hyper}

\date{\vspace{-5em}}
\usepackage[affil-it]{authblk} 
\usepackage{etoolbox}
\usepackage{lmodern}
\makeatletter
\patchcmd{\@maketitle}{\LARGE \@title}{\fontsize{16}{19.2}\selectfont\@title}{}{}
\makeatother

\title{\vspace{-5em} Identifying the minimal sets of distance restraints for FRET-assisted protein structural modeling}

\author[1,2]{Zhuoyi Liu}
\author[2,3]{Alex T. Grigas}
\author[2,3]{Jacob Sumner}
\author[4]{Edward Knab}
\author[4]{Caitlin M. Davis}
\author[1,2,3,5,6]{Corey S. O'Hern}
\affil[1]{Department of Mechanical Engineering and Materials Science, Yale University, New Haven, Connecticut, 06520, USA}
\affil[2]{Integrated Graduate Program in Physical and Engineering Biology, Yale University, New Haven, Connecticut, 06520, USA}
\affil[3]{Graduate Program in Computational Biology and Bioinformatics, Yale University, New Haven, Connecticut, 06520, USA}
\affil[4]{Department of Chemistry, Yale University, New Haven, Connecticut, 06520, USA}
\affil[5]{Department of Physics, Yale University, New Haven, Connecticut, 06520, USA}
\affil[6]{Department of Applied Physics, Yale University, New Haven, Connecticut, 06520, USA}

\begin{document}

\maketitle

\textbf{}

\textbf{Manuscript Pages: 17}

\textbf{Total Manuscript Figures:} 6

\textbf{Total Manuscript Tables:} 1

\textbf{Supporting Information Pages: 11}

\textbf{Total Supporting Information Figures/Tables: 10/0}

\textbf{}

\textbf{Abstract:} Proteins naturally occur in crowded cellular environments and interact with other proteins, nucleic acids, and organelles. Since most previous experimental protein structure determination techniques require that proteins occur in idealized, non-physiological environments, the effects of realistic cellular environments on protein structure are largely unexplored. Recently, F\"{o}rster resonance energy transfer (FRET) has been shown to be an effective experimental method for investigating protein structure {\it in vivo}. Inter-residue distances measured {\it in vivo} can be incorporated as restraints in molecular dynamics (MD) simulations to model protein structural dynamics {\it in vivo}.  Since most FRET studies only obtain inter-residue separations for a small number of amino acid pairs, it is important to determine the minimum number of restraints in the MD simulations that are required to achieve a given root-mean-square deviation (RMSD) from the experimental structural ensemble. Further, what is the optimal method for selecting these inter-residue restraints? Here, we implement several methods for selecting the most important FRET pairs and determine the number of pairs $N_{r}$ that are needed to induce conformational changes in proteins between two experimentally determined structures. We find that enforcing only a small fraction of restraints, $N_{r}/N \lesssim 0.08$, where $N$ is the number of amino acids, can induce the conformational changes. These results establish the efficacy of FRET-assisted MD simulations for atomic scale structural modeling of proteins {\it in vivo}.   

\textbf{Significance:} Determining protein structure {\it in vivo} is essential for understanding protein function.  Most protein structures have been studied in non-physiological conditions using x-ray crystallography, NMR spectroscopy, and cryo-electron microscopy. Thus, we do not know whether the cellular environment significantly affects protein structure. We emphasize the benefits of FRET-assisted molecular dynamics simulations in characterizing protein structure {\it in vivo} at the atomic scale. We identify the minimum number of FRET pairs that can induce conformational changes in several proteins, including one that has been characterized using in-cell NMR.


\newpage

\section{\label{sec:intro}Introduction}

Knowing the three-dimensional structure of proteins enables us to understand the biophysical mechanisms that control protein function, protein-protein interactions, and cell signaling. Nearly all protein structures that have been determined experimentally to date have been characterized under idealized, non-physiological conditions. For example, proteins have been crystallized into non-native, solid phases for x-ray scattering experiments~\cite{smyth2000x}, and proteins have been dissolved into dilute, non-physiological buffers for NMR spectroscopy or cryo-electron microscopy~\cite{williamson1985solution,carroni2016cryo}. However, proteins carry out their functions in cellular environments that are significantly different from these {\it in vitro} conditions. The cellular environment is crowded with a non-solvent packing fraction of $0.3$-$0.4$ that includes nucleic acids, carbohydrates, lipids, organelles, and other components~\cite{fulton1982crowded,zimmerman1991estimation,ellis2001macromolecular}. This environment impacts the physical properties of proteins, including the protein's radius of gyration, melting temperature, and rotational diffusion coefficient~\cite{ebbinghaus2010protein,davis2020vitro,leeb2020diffusive,wang2018cell}. Currently, there are over $200$,$000$ {\it in vitro} protein structures deposited in the Protein Data Bank (PDB)~\cite{berman2000protein}, as well as more than $10^6$ computational models predicted by 
 AlphaFold and RosettaFold~\cite{jumper2021highly,humphreys2021computed}. To date, {\it in vivo} structures have been obtained for only three proteins (TTHA1718, GB1, and ubiquitin) using in-cell NMR~\cite{sakakibara2009protein,tanaka2019high,gerez2022protein}. In-cell NMR is not widely used since it is difficult to distinguish the isotopic labeled target protein from its environment, which leads to a low signal-to-noise ratio and sensitivity~\cite{ikeya2019protein,serber2001cell,luchinat2023cell}. Another experimental method for solving protein structure {\it in vivo} is cryo-electron tomography~\cite{cheng2023determining}, however, this technique requires proteins to be confined to a thickness less than $500$~nm, which limits this method to only a subset of cell types and membrane-associated proteins~\cite{dunstone2017cryo,hylton2021challenges}. In addition, there have been numerous computational studies of proteins {\it in vivo}. All-atom molecular dynamics (MD) simulations have investigated protein structure in the presence of nearly all components of the cell cytoplasm~\cite{rickard2020crowding,stevens2023molecular}. However, current force fields have been calibrated to {\it in vitro} protein structures and we do not have accurate potentials for interactions between proteins and nucleic acids, ribosomes, and organelles~\cite{tucker2022development,love2023assessing}. Thus, we do not yet have a quantitative, atomistic-level understanding of protein structure in cells.

F\"{o}rster resonance energy transfer (FRET) can be used to determine the separations between donor and acceptor chromophores attached to a pair of amino acids in a given protein by measuring their energy transfer efficiency. The distribution of separations between the two amino acids can then be deduced by calculating the configuration space volumes sampled by the donor and acceptor chromophores~\cite{kalinin2012toolkit,dimura2016quantitative,klose2021resolving}. FRET pair labeling is highly specific and offers high accuracy for the inter-residue separations with uncertainties less than 2-4 \AA~~\cite{hellenkamp2018precision,agam2023reliability}. Additionally, FRET-labeled proteins can be directly expressed or injected into cells, enabling single molecule measurements of protein structure, dynamics, and stability {\it in vivo}~\cite{feng2019quantifying,ebbinghaus2010protein,davis2018non}. However, most FRET experimental studies only obtain inter-residue separations for a small number of amino acid pairs in a given protein~\cite{ebbinghaus2010protein,wang2018cell,davis2018non}. To investigate the atomic scale structure of proteins {\it in vivo}, inter-residue separations obtained from FRET experiments can be incorporated into molecular dynamics (MD) simulations. Current all-atom MD simulations of proteins using {\it in vitro} solution conditions have been shown to sample {\it in vitro} protein structures obtained from x-ray crystallography and NMR spectroscopy~\cite{lindorff2011fast,lindorff2012systematic}. By including a sufficient number of inter-residue restraints from FRET studies of proteins {\it in vivo} into the MD force fields that were developed for {\it in vitro} solution conditions, it may be possible to sample the {\it in vivo} structural ensemble of proteins.

Several key questions must be answered before FRET-assisted structural modeling of protein structure {\it in vivo} can be employed. In particular, given the large number of possible FRET pairs, what is the minimum number of amino acid pairs for which we need distance restraints to achieve a given accuracy for the root-mean-square deviations (RMSD) between the C$_{\alpha}$ positions in the restrained simulations and those in the experimental structures? However, we do not yet have access to high-resolution {\it in vivo} protein structures. Thus, we will first develop the methodology for carrying out restrained MD simulations for protein structural modeling using proteins that are found in multiple conformational states {\it in vitro}. The hypothesis is that the method that we use to induce conformational changes {\it in vitro} can also be used to study conformational changes in {\it in vivo} environments. The initial conformational state will be metastable in the MD force field without restraints, and we will induce a conformational change in the protein by incorporating restraints between amino acid pairs that are satisfied in the target state. To our knowledge, there have only been a few studies aimed at identifying the most important restraints for efficiently moving to the conformational ensemble of the target state~\cite{dimura2020automated}. For example, currently we do not know the minimal number of restraints and which restraints are necessary to achieve a given RMSD in the C$_{\alpha}$ positions from the target state, and how random selection of given number of restraints compares to other methods for selecting restraints. To connect the {\it in vitro} results to those for {\it in vivo} conditions, we will also study one of the few proteins whose structure has been solved {\it in vivo} using in-cell NMR spectroscopy. 

Here, we select four proteins that each can take on two, distinct conformational states to develop the restrained MD simulation methodology: T4 Lysozyme (172L/1L69), Phosphoglycerate Kinase (2XE6/2Y3I), Adenylate Kinase (4AKE/1AKE), and Tick
Carboxypeptidase Inhibitor (1ZLI/2JTO), where the first and second PDBIDs indicate the initial and target structures, respectively. We initialize the MD simulations with the crystal structure of the initial state, add a given number of C$_{\alpha}$ distance restraints, and calculate the C$_{\alpha}$ RMSD relative to the target structure. For the first three proteins (T4 Lysozyme (T4L), Phosphoglycerate Kinase (PGK), and Adenylate Kinase (AK)) that are metastable in the force field, we test four methods (normal mode analysis, the largest C$_{\alpha}$ separation method, the largest change in pairwise separation method, and linear discriminant analysis) for selecting the restraints and compare the C$_{\alpha}$ RMSD to the target structure to that obtained from random selection of the restraints. We also vary the number of restraints to determine the minimum number of restraints needed to achieve a given C$_{\alpha}$ RMSD from the target. Two of the methods (normal mode analysis and the largest C$_{\alpha}$ separation method) do not use information about the target structure, whereas the other two methods (the largest change in pairwise separation method and linear discriminant analysis) compare the initial and target structures to identify the most important restraints. We find that for T4L, PGK, and AK, which take on two distinct conformational states {\it in vitro}, we can induce the conformational changes using only a small fraction of restraints, $N_{r}/N \lesssim 0.02$, where $N$ is the number of amino acids in the protein. For the proteins that we considered, this result corresponds to $1$-$5$ restraints, which is a number that can readily be achieved in FRET experiments. In addition, we studied one of the few proteins that has been characterized using in-cell NMR spectroscopy: the B1 domain of protein G (GB1). We need a slightly larger fraction of restraints, $N_r/N \sim 0.08$ (or $5$ restraints), to change the protein conformation from the initial {\it in vitro} structure (2N9K) to the in-cell NMR structure (7QTS). Using our methods to induce conformational changes from the bound to the unbound conformations of Tick Carboxypeptidase Inhibitor (TCI) and from the {\it in vivo} to the {\it in vitro} structures of GB1, we show that the fraction of restraints required to induce conformational changes depends on the stability of the target state in the force field. In general, the largest change in pairwise separation method, which has information about the target structure, yields the lowest values for the C$_{\alpha}$ RMSD for a given $N_r$. If the restraint selection method does not have information about the target structure, the C$_{\alpha}$ RMSD is still lower than that for random selection. These results establish the feasibility of FRET-assisted structural modeling of proteins {\it in vivo} using restrained MD simulations, but also emphasize the need for improved force fields for MD simulations of proteins {\it in vivo}.

\section{Materials and Methods}

\subsection*{Selected Proteins}

We identified three proteins from the Protein Data Bank (PDB) that can be crystallized into two distinct conformational states {\it in vitro} and can be used to develop the restrained MD simulation methodology: T4 Lysozyme (T4L), Phosphoglycerate Kinase (PGK), and Adenylate Kinase (AK). (See Table~\ref{table:1}.)~\cite{zhang1992multiple,zhang1995protein,zerrad2011spring,lallemand2011interaction,muller1992structure,muller1996adenylate}. These proteins have been studied extensively in the context of protein conformational changes~\cite{flores2006database}. While the selected targets are {\it in vitro} structures, recent studies suggest that the differences between the {\it in vivo} and {\it in vitro} structures for PGK are largely caused by the hinge motion of the two domains that occurs between the initial and target {\it in vitro} structures~\cite{davis2018non,davis2020vitro}. Previous studies have also suggested that the crowded cellular environment can stabilize the target {\it in vitro} structure for AK~\cite{li2014combined}. For the restrained MD simulations, we require that the initial structure is stable and the target structure is at least metastable over long time scales in MD simulations, which ensures that the unrestrained dynamics does not induce the transition from the initial state to the target. (See Fig. S1 in Supporting Information.) We also consider one protein that has both an {\it in vitro} and in-cell NMR structure, the B1 domain of protein G (GB1)~\cite{ikeya2016improved,gerez2022protein}. We use the first model from the {\it in vitro} NMR bundle as the initial structure and the first model of the in-cell NMR bundle as the target structure. (See Fig. S6 in Supporting Information.) We find similar results for the restrained and unrestrained MD simulations when we initialize the system using the other NMR models. To investigate the dependence of the restraint selection method on the stability of the target structure, we study heterodimer, Tick Carboxypeptidase Inhibitor (TCI)~\cite{pantoja2008nmr,arolas2005three}, which is metastable in its bound conformation, but unstable in its unbound conformation. (See Fig. S7 in Supporting Information.)

\begin{table}[h]
\centering
\footnotesize
\begin{tabular}{c c c c c c}
\hline
Protein & initial structure & target structure & $N$ & C$_{\alpha}$ ${\rm RMSD}$ (\AA) & $T_m$ (K)\\\hline
T4 Lysozyme (T4L) & 172L & 1L69 & 162 & 3.95 & 344~\cite{baase2010lessons}\\
Phosphoglycerate Kinase (PGK) & 2XE6 & 2X15 & 413 & 4.08 & 327~\cite{fiorillo2018phosphoglycerate}\\
Adenylate Kinase (AK) & 4AKE & 1AKE & 214 & 5.96 & 333~\cite{chang2021rational}\\
B1 domain of protein G (GB1) & 2N9K & 7QTS & 57 & 2.86 & 354~\cite{campos2002placement}\\
Tick Carboxypeptidase Inhibitor (TCI) & 1ZLI & 2JTO & 77 & 3.82 & $>$343~\cite{arolas2006carboxypeptidase}\\
\hline
\end{tabular}
\caption{We list the selected proteins including their PDBIDs for the initial and target structures, number of amino acids $N$, the C$_{\alpha}$ RMSD (Eq.~\ref{equation:2}) between the initial and target structures, and the melting temperature $T_m$. For GB1, the C$_{\alpha}$ RMSD is calculated between the first model of the {\it in vitro} NMR bundle and the first model of the in-cell NMR bundle.}
\label{table:1}
\end{table}

\subsection*{C$_{\alpha}$ RMSD}

To compare the protein structures from the restrained MD simulations (i.e. structure $S_i$) and the target structure (i.e. structure $S_j$), we 
define the C$_{\alpha}$ separation vector for the $\beta$th amino acid, 
\begin{equation}
{\vec \Delta}(S_{i},S_{j};\beta) = \vec{r}_{i,\beta} - \vec{r}_{j,\beta},
\end{equation}
where $\vec{r}_{i,\beta}$ is the position of the $C_{\alpha}$ atom on amino acid $\beta$ in structure $S_{i}$. We define the root-mean-square deviation in the C$_{\alpha}$ positions between two structures $S_i$ and $S_j$ as
\begin{equation}
{\rm RMSD}(S_{i},S_{j}) = \sqrt{\frac{1}{N}\sum_{\beta=1}^{N}\Delta^2(S_{i},S_{j},\beta)}.
\label{equation:2}
\end{equation}
When comparing two structures $S_i$ and $S_j$, we typically align them (i.e. rotate one of them) to achieve the minimum value of the C$_{\alpha}$ ${\rm RMSD}(S_i,S_j)$ for a given $S_i$ and $S_j$. 
We can also calculate the C$_{\alpha}$ RMSD between the restrained and target structures averaged over an ensemble of restrained structures for each amino acid $\beta$: 
\begin{equation}
{\rm RMSD}(\{S_i\}, S_{j}, \beta) = \sqrt{\frac{1}{N_{s}}\sum_{i=1}^{N_s}\Delta^2(S_{i},S_{j},\beta)},
\label{equation:3}
\end{equation}
where $N_s$ is the number of protein structures in the restrained ensemble $\{S_i\}$ and $S_{j}$ represents the target structure. 

\subsection*{Restraint selection methods}

Below, we describe the methods that we employ for selecting the C$_{\alpha}$ distance restraints. Linear spring restraints will be added to MD simulations of the initial protein structure, where the rest lengths are obtained from the target structure, as discussed below. Three of the restraint selection methods, random selection, normal mode analysis, and the largest C$_{\alpha}$ separation method, do not use information about the target structure. Two additional methods, the largest change in pairwise separation method and linear discriminant analysis, compare the initial {\it and} target structure to identify the most effective restraints.

\subsubsection*{a) Random selection}

In a protein with $N$ amino acids, there are $N_p=N(N-1)/2$ distinct amino acid pairs. As shown in Fig.~\ref{fig:1}a, the pair separations between C$_{\alpha}$ atoms can be represented using a symmetric distance matrix $R_{\beta \delta}$, where $\beta$,$\delta=1,\ldots,N$. To establish a baseline for the performance of the restrained MD simulations, we will first consider random selection of the restraints. In this approach, we exclude amino acid pairs that are too close ($R_{\beta \delta}<R_g$, where $R_g$ is the radius of gyration), and pairs for which at least one amino acid is buried with relative solvent accessible surface area ${\rm rSASA} < 0.1$ since such pairs will preclude FRET measurements~\cite{grigas2022core,richards1974interpretation}, which reduces the pool of restraints to approximately $N_p/3$ amino acid pairs. For each number of restraints $N_{r} \ll N_p$ we consider, we select $100$ sets of $N_r$ restraints randomly from the pool of allowed pairs.

\subsubsection*{b) Normal mode analysis}

\begin{figure}[ht]
\centering
\includegraphics[width=1\linewidth]{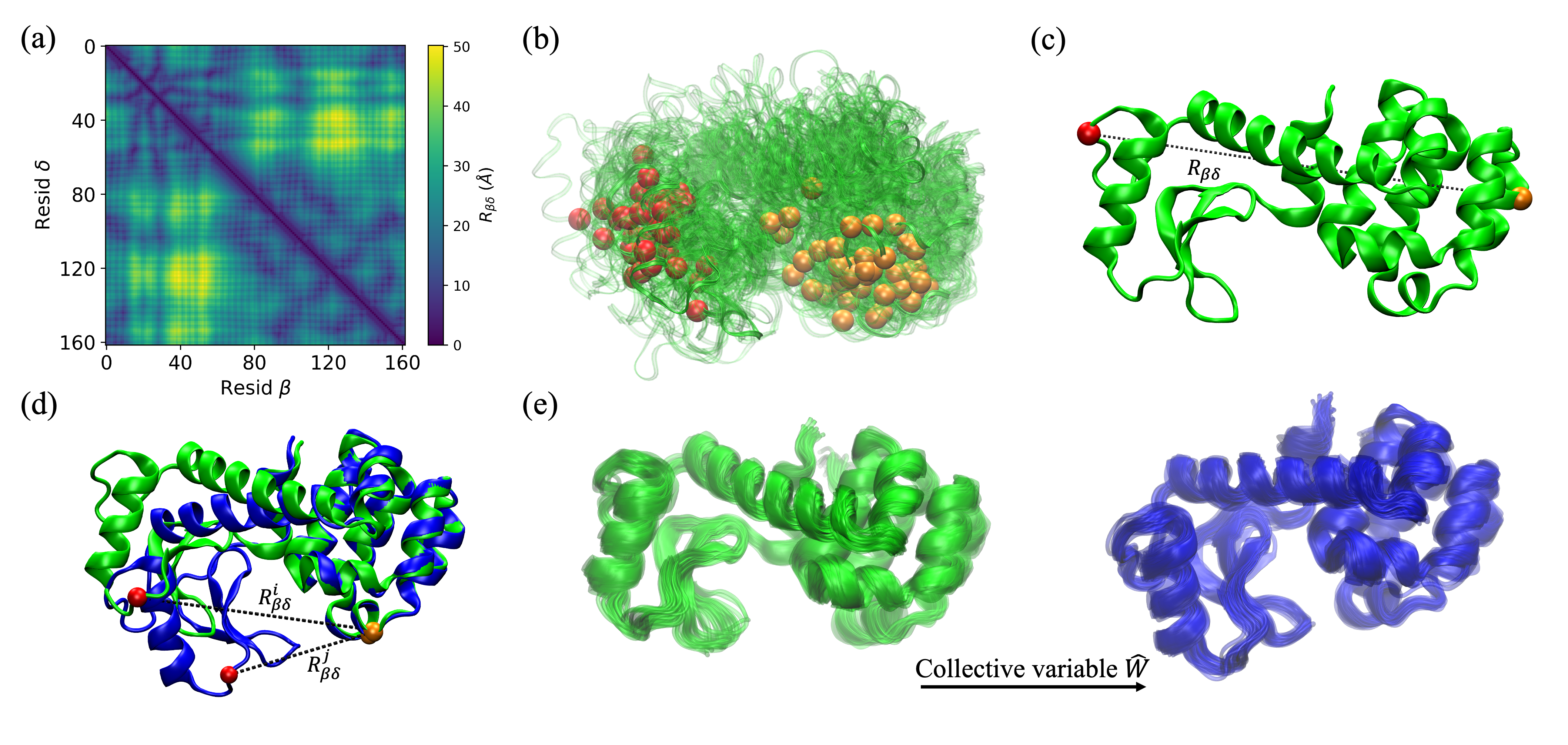}
\caption{Illustration of several distance restraint selection methods, using T4L as the example protein. (a) The symmetric distance matrix $R_{\beta \delta}$, where the color gradient from dark to light indicates increasing C$_{\alpha}$ separations between amino acids $\beta$ and $\delta$. (b) For the normal mode analysis method, we generate an ensemble of $10^4$ structures ($100$ structures are shown in green) that have been displaced from the initial structure along a random superposition of normal modes corresponding to the ten lowest frequencies. A single amino acid pair that minimizes the $C_{\alpha}$ RMSD among structures in the ensemble is highlighted in red and orange in each structure. (c) We identify the amino acid pairs with the largest C$_{\alpha}$ separations $R_{\beta \delta}$ in the initial (green) structure. The pair with the maximum $R_{\beta \delta}$ is highlighted in red and orange. (d) We can also select restraints by identifying the largest changes in the pairwise C$_{\alpha}$ separations between the initial (green) and target (blue) structures. The C$_{\alpha}$ atoms of the pair with the maximum $\Delta R_{\beta \delta}(S_i,S_j)$ are shown in red and orange. (e) $20$ conformations from unrestrained MD simulations starting from the initial and target structures are shown in green and blue, respectively. Using linear discriminant analysis, we identify the collective variable ${\hat W}$ that maximizes the cross-correlation of the two ensembles.}
\label{fig:1}
\end{figure}

For this method, we assume that the normal modes of vibration of the initial structure provide information about how the protein transitions from the initial to the target structure. We follow the methodology used in other recent work aimed at selecting the most effective FRET pairs for restrained MD simulations of proteins~\cite{dimura2020automated}. First, the method constructs a coarse-grained elastic network from the atomic positions of the initial protein structure and calculates the normal modes of the network~\cite{ahmed2006multiscale}. Then, the initial structure is displaced by a linear combination of the ten lowest frequency modes with amplitudes that are inversely proportional to the frequency and have proportionality constants $-1 < \xi < 1$ that are chosen randomly. The normal modes are then recalculated on the displaced structure and the structure is perturbed again using a linear combination of the ten lowest frequency modes with amplitudes chosen as before. This process of successive displacements along the lowest frequency normal modes is continued until $10^3$ structures are obtained~\cite{kruger2012nmsim} and then repeated ten times with independently generated random mode amplitudes to yield a total of ${\cal N} =10^4$ structures. $100$ of the structures for T4L are shown in Fig.~\ref{fig:1}b.

To identify the amino acid pairs that should be restrained, we seek to minimize the RMSD of the positions of the C$_\alpha$ atoms (Eq.~\ref{equation:2}) among the protein structures in the ensemble, where the weights for each structure in the ensemble are controlled by the size of the fluctuations in the amino acid pair separations among structures. Deviations in the pair separations between two structures $S_{i}$ and $S_{j}$ can be quantified using
\begin{equation}
\chi^2(S_i, S_j) = \sum_{ \{ (\beta,\delta) \}}\left(\frac{\Delta R_{\beta \delta} (S_i,S_j)}{R_{\beta \delta}^i+R^j_{\beta \delta}}\right)^2,
\end{equation}
where $\Delta R_{\beta \delta} (S_i,S_j) = |R_{\beta \delta}^i-R^j_{\beta \delta}|$, $\{(\beta,\delta)\}$ is a given set of amino acid pairs, and $R_{\beta \delta}^i$ is the distance between C$_{\alpha}$ atoms on amino acids $\beta$ and $\delta$ on structure $S_i$. To estimate the probability of observing a mean-square deviation in the pair separations larger than $\chi^2(S_i, S_j)$, we calculate 
\begin{equation}
P(S_i, S_j) = \int_{\chi^2(S_i, S_j)}^{\infty} f(N_m,\chi^2) \,d\chi^2,
\end{equation}
where $N_m$ is the number of amino acid pairs in the set $\{(\beta,\delta)\}$ and $f(N_m,\chi^2)$ is the chi-squared distribution with $N_m$ degrees of freedom. To quantify the average C$_{\alpha}$ RMSD over the ensemble, we calculate
\begin{equation}
\langle {\rm RMSD} \rangle = \frac{1}{\cal N}\sum_{i=1}^{\cal N}\frac{\sum_{j=1}^{\cal N}P(S_i, S_j){\rm RMSD}(S_i,S_j)}{\sum_{j=1}^{{\cal N}} P(S_i, S_j)}.
\end{equation}
To minimize $\langle {\rm RMSD} \rangle$, the structures for which the selected amino acid pairs have large deviations in their pair separations (i.e. small $P(S_i,S_j)$) should possess large C$_\alpha$ $\rm RMSD$, and the structures for which the selected amino acid pairs have small deviations (i.e. large $P(S_i,S_j)$) should possess small C$_\alpha$ $\rm RMSD$. To select the set of amino acid pairs that minimize $\langle {\rm RMSD} \rangle$, we add one pair to the set of optimal pairs at a time. We start with identifying the single pair $(\beta^{*},\delta^{*})$ that minimizes $\langle {\rm RMSD} \rangle$ and use it as the restraint for $N_r=1$. We then consider two possible pairs, but fix one of the pairs to be $(\beta^{*},\delta^{*})$ and find the new pair $(\beta^{**},\delta^{**})$ in the set $\{(\beta^{*},\delta^{*}),(\beta^{**},\delta^{**})\}$ that minimizes $\langle {\rm RMSD} \rangle$. We use these two pairs as the set of restraints for $N_r=2$. This process continues until we have $N_r=1,\ldots,N_{\rm max}$, where $N_{\rm max}/N < 0.08$ for all proteins considered.

\subsubsection*{c) Identifying the largest C$_{\alpha}$ separations in the initial structures}

For this method (i.e. the largest C$_{\alpha}$ separation method) for selecting important restraints, we identify the amino acid pairs with the largest C$_{\alpha}$ separations, or the maximum $R_{\beta \delta}$ over all pairs $\{(\beta,\delta)\}$ in the initial protein structure. (For example, in Fig.~\ref{fig:1}c, we show the amino acid pair with the largest C$_{\alpha}$ separation in the x-ray crystal structure for T4L (172L).) After identifying the pair $(\beta,\delta)$ with the largest C$_{\alpha}$ separation, we find the pair $(\beta^{'},\delta^{'})$ with the second largest C$_{\alpha}$ separation. This process is then continued to find a set of pairs with the largest C$_{\alpha}$ separations. However, we seek to identify a minimal set of amino acid pairs without redundant structural information. Thus, we carried out unrestrained MD simulations starting from the initial structure and determined the Pearson correlation between the pairwise C$_{\alpha}$ separations. In particular, we include $(\beta^{'},\delta^{'})$ in the pool of selected pairs if the Pearson correlation $\rho$ between $R_{\beta \delta}$ and $R_{\beta^{'} \delta^{'}}$ satisfies $|\rho| < 0.9$. We also require that the C$_{\alpha}$ atoms of the new pair are not already in the pool of selected restraints. Thus, we first add the pair $(\beta,\delta)$ with the largest C$_{\alpha}$ separation to the pool of restraints ($N_r=1$). We add $(\beta',\delta')$ with the second largest C$_{\alpha}$ separation to the pool of restraints ($N_r=2$) as long as it is uncorrelated with $(\beta,\delta)$ and does not include C$_{\alpha}$ atom $\beta$ or $\delta$.  If so, we consider the pair $(\beta'',\delta'')$ with the next largest separation. We follow this process until we have sets of restraints with $N_r=1,\ldots,5$.

\subsubsection*{d) Identifying the largest change in pairwise C$_{\alpha}$ separations between the initial and target structures}

For this method (i.e. the largest change in pairwise separation method), we assume that the target structure is known. We identify the amino acid pair with the largest change in pairwise C$_{\alpha}$ separations between the initial and target structures $S_i$ and $S_j$ (i.e. the maximum $\Delta R_{\beta \delta} (S_i,S_j)$). (For example, in Fig.~\ref{fig:1}d, we show the amino acid pair with the largest change in C$_{\alpha}$ separations between the initial (172L) and target (1L69) x-ray crystal structures for T4L. Note that the amino acids with the largest change in pairwise separation are not the same as those with the largest C$_{\alpha}$ separation.) To identify a set of non-redundant restraints, we successively implement new restraints in MD simulations for pairs with the largest deviation in the pair separation from the target. In particular, assume that pair $(\beta, \delta)$ has the largest $\Delta R_{\beta \delta} (S_i,S_j)$ in the unrestrained MD simulations starting from the initial structure.  We then carry out MD simulations with $R_{\beta \delta}$ restrained to the target value. We identify the pair $(\beta^{'}, \delta^{'})$ with the largest $\Delta R_{\beta^{'} \delta^{'}} (S_i,S_j)$ and carry out restrained MD restraints enforcing the target values for both $R_{\beta \delta}$ and $R_{\beta^{'} \delta^{'}}$. We also require that the C$_{\alpha}$ atoms of the new pair are not the same as any of the atoms in the current pool of restraints. We use $(\beta, \delta)$ as the restraint for $N_r=1$ and use $(\beta, \delta)$ with $(\beta^{'},\delta^{'})$ as the set of restraints for $N_r=2$. This process continues until we have $N_r=1,\ldots,N_{\rm max}$, where $N_{\rm max}/N <0.08$ for all proteins we considered.

\subsubsection*{e) Linear discriminant analysis}

For the linear discriminant analysis method of selecting restraints, we also assume that the target structure is known. We seek to identify the pairwise separation between C$_{\alpha}$ atoms that can serve as a collective variable enabling the protein to move from the initial to the target structure~\cite{sittel2018perspective}. The inspiration for this method is the linear discriminant analysis method that was used to identify the collective variables for small molecule conformational changes~\cite{mendels2022collective}. Here, we apply the method in the context of large conformational changes in proteins. 

Linear discriminant analysis requires a distribution of protein structures, not a single structure. Thus, we first carried out short $20$ ns unrestrained MD simulations starting from both the initial and target structures to sample an ensemble of structures near the initial and target structures. In Fig.~\ref{fig:1}e, we show these two distributions of structures for T4L as an example. In this case, we aim to find the collective variable that tracks the hinge closure motion of the two domains. Each protein conformation can be described by the set of all pairwise separations between C$_{\alpha}$ atoms,
\begin{equation}
{\vec X} = (R_{\beta \delta}, R_{\beta^{'} \delta^{'}}, ...)^T,
\end{equation}
where the length of the vector is the number of distinct amino acid pairs $N_p$ minus the ones that are too close together ($R_{\beta \delta}<R_g$) and the ones where at least one amino acid is buried. We can describe the distribution of protein conformations sampled near the initial and target structures as $\rho_A({\vec X}_{A})$ and $\rho_A({\vec X}_{B})$, respectively. We aim to find the direction ${\hat W}$ in the space of C$_{\alpha}$ separations such that its projection onto the cross-correlation matrix (between initial and target distributions) is maximized, while its projection onto the covariance matrix for the initial or target distributions is minimized. To determine ${\hat W}$, we first calculate the mean separations $\langle {\vec X}_A \rangle$ and $\langle {\vec X}_B \rangle$ from the distributions of the initial and target structures. Second, we define the (cross-correlation) scattering matrix 
\begin{equation}
S_{b} = (\langle {\vec X}_A \rangle - \langle {\vec X}_B \rangle)(\langle {\vec X}_A \rangle - \langle {\vec X}_B \rangle)^T,
\end{equation}
which quantifies the covariance of the initial distribution with the target distribution, and define the projection of ${\hat W}$ onto $S_b$ as ${\hat W}^TS_{b}{\hat W}$. We also quantify the covariance of the initial and target distributions, separately:
\begin{equation}
\Sigma_{A,B}  = (X_{A,B} - \langle X_{A,B} \rangle)(X_{A,B} - \langle X_{A,B} \rangle)^T.
\end{equation}
We can then calculate the (direct correlation) scattering matrix,
\begin{equation}
S_{w} = \Sigma_{A} + \Sigma_{B},
\end{equation}
with projection ${\hat W}^TS_{w}{\hat W}$. To find the direction ${\hat W}$ onto which the projections of the two distributions $\rho_A({\vec X}_A)$ and $\rho_B({\vec X}_B)$ are best separated, we can maximize the Rayleigh ratio:
\begin{equation}
\mathcal{R}(\hat W) = \frac{{\hat W}^TS_{b}{\hat W}}{{\hat W}^TS_{w}{\hat W}},
\end{equation}
which is equivalent to solving for the eigenvector ${\hat W}_{\lambda}$ associated with the largest eigenvalue $\lambda$ of $S_{w}^{-1}S_{b}$:
\begin{equation}
S_{w}^{-1}S_{b}{\hat W}_{\lambda} = \lambda{\hat W}_{\lambda}.
\end{equation}
We identify the most important pairwise distances as those that have the largest weights in ${\hat W}_{\lambda}$. For $N_r=1$, we choose the amino acid pair $(\beta,\delta)$ with the largest weight. For $N_r=2$, we choose the two amino acid pairs $(\beta,\delta)$ and $(\beta^{'}, \delta^{'})$ with the two largest weights, and so on.  

\subsection*{Restrained MD simulations}

To assess the performance of the FRET pair selection methods, we carry out all-atom MD simulations starting from the initial structure and incorporating the C$_{\alpha}$-C$_{\alpha}$ separations of the selected pairs from the target structure as the equilibrium lengths of the linear spring restraints. We then monitor the RMSD of the C$_{\alpha}$ atom positions (Eq.~\ref{equation:2}) from the target structure as a function of time. Unrestrained and restrained MD simulations were performed using the AMBER99SB-ILDN force field~\cite{best2009optimized,lindorff2010improved} in the GROMACS software package~\cite{abraham2015gromacs}. The MD simulations were carried out in a periodic dodecahedron-shaped box that is sufficiently large such that the protein surface is at least $20$\AA~from the box edges. The simulation box was solvated with water molecules modeled using TIP3P at neutral pH and $0.15$M NaCl~\cite{jorgensen1981transferable,mark2001structure}. Short-range van der Waals and screened Coulomb interactions were truncated at $1.2$ nm, while long-ranged electrostatic interactions were tabulated using the Particle Mesh Ewald summation method. The LINCS algorithm was used to constrain the bond lengths. We performed two energy minimization runs to first relax the protein and then the water molecules and the protein together using the steepest decent algorithm until the maximum net force magnitude on an atom is smaller than $500$ $\text{kJmol}^{-1}\text{nm}^{-1}$. We perform NVT simulations of the system for $20$ ns at temperature $T=300$K using a velocity rescaling thermostat for sampling the canonical ensemble~\cite{bussi2007canonical}. The equations of motion for the atomic coordinates and velocities are integrated using a leapfrog algorithm with a $1$~fs time step. 

For the restrained simulations, we employ a linear spring potential for each restrained amino acid pair $(\beta,\delta)$:
\begin{equation}
E_{r}(R_{\beta \delta}) = \frac{k_r}{2}(R_{\beta \delta}-R^{0}_{\beta \delta})^2,
\end{equation}
where $R^{0}_{\beta \delta}$ is the C$_{\alpha}$-C$_{\alpha}$ separation for the $(\beta,\delta)$ pair in the target structure and the spring constant $k_{r}=5000$ $\text{kJmol}^{-1}\text{nm}^{-2}$ is chosen so that the averaged root-mean-square deviation between $R_{\beta \delta}$ and $R^{0}_{\beta \delta}$ is $< 0.2$\AA. (See Fig. S2 in Supporting Information.) 

For the unrestrained MD simulations for each protein, we ran $N_s=100$ simulations for 20 ns starting from the initial structure, but with different initial velocities for each atom randomly selected from a Maxwell-Boltzmann distribution at $T=300$K. Using these simulations, we can calculate the average C$_{\alpha}$ RMSD between the initial structures and the target structure, which serves as a baseline for comparison to the results for the restrained MD simulations. For the random restraint selection method at each $N_r$, we randomly select $N_s=100$ sets of $N_r$ amino acid pairs and run one restrained MD simulation (for 20 ns) for each set of restraint pairs. For all other restraint selection methods, we carry out $N_s=50$ simulations for each set of $N_r$ restraints, but with different random initial velocities. The number of samples $N_s$ is chosen so that average C$_\alpha$ $\rm RMSD$ converges at large $N_s$ as shown in Fig. S3 in the Supporting Information. For each $N_r$ and for each restraint selection method, we average the C$_\alpha$ $\rm RMSD$ over all $N_s$ simulations. For the restrained MD simulations of the {\it in vitro} protein structures, the C$_{\alpha}$ RMSD converges within the first $5$ ns, as shown in Fig. S4 in the Supporting Information. Therefore, we only use the data generated from $5$ to $20$ ns for the calculations of the C$_{\alpha}$ $\rm RMSD$. For the {\it in vivo} target GB1, we extended the MD simulations to $50$ ns and use the time points from $20$ to $50$ ns to calculate the C$_{\alpha}$ RMSD. (See Fig. S6c in Supporting Information.)

To determine the ability of the restraints to move the protein conformation from the initial to the target structure, we need to measure a reference C$_{\alpha}$ RMSD (i.e. RMSD$_r$) that quantifies thermal fluctuations around the target structure. For the proteins T4L, PGK, and AK, we determine RMSD$_r$ by running unrestrained MD simulations starting from the target structure. We set RMSD$_r$ = ${\rm RMSD}(S_i,S_j)$, where $S_i$ is the central structure from the unrestrained MD simulations starting from the target structure, and $S_j$ is the target structure. We find the central structure by identifying the structure that has the largest number of neighboring structures in the ensemble using a cutoff C$_\alpha$ RMSD of 1\AA. If the average C$_\alpha$ RMSD obtained from the restrained MD simulations is below RMSD$_r$, the restraints were successful in moving the initial structure to the target structure since the differences are comparable to thermal fluctuations observed near the target structure. For the {\it in vitro} and in-cell NMR structures of GB1 and the unbound structure of TCI, we set RMSD$_r = N_m^{-1} \sum_{i>j} {\rm RMSD}(S_i,S_j)$, where $S_i$ and $S_j$ are distinct models in the in-cell NMR bundle and $N_m$ is the number of distinct pairs of models. 

\section{Results}
\label{results}

\begin{figure}[ht]
\centering
\includegraphics[width=1\linewidth]{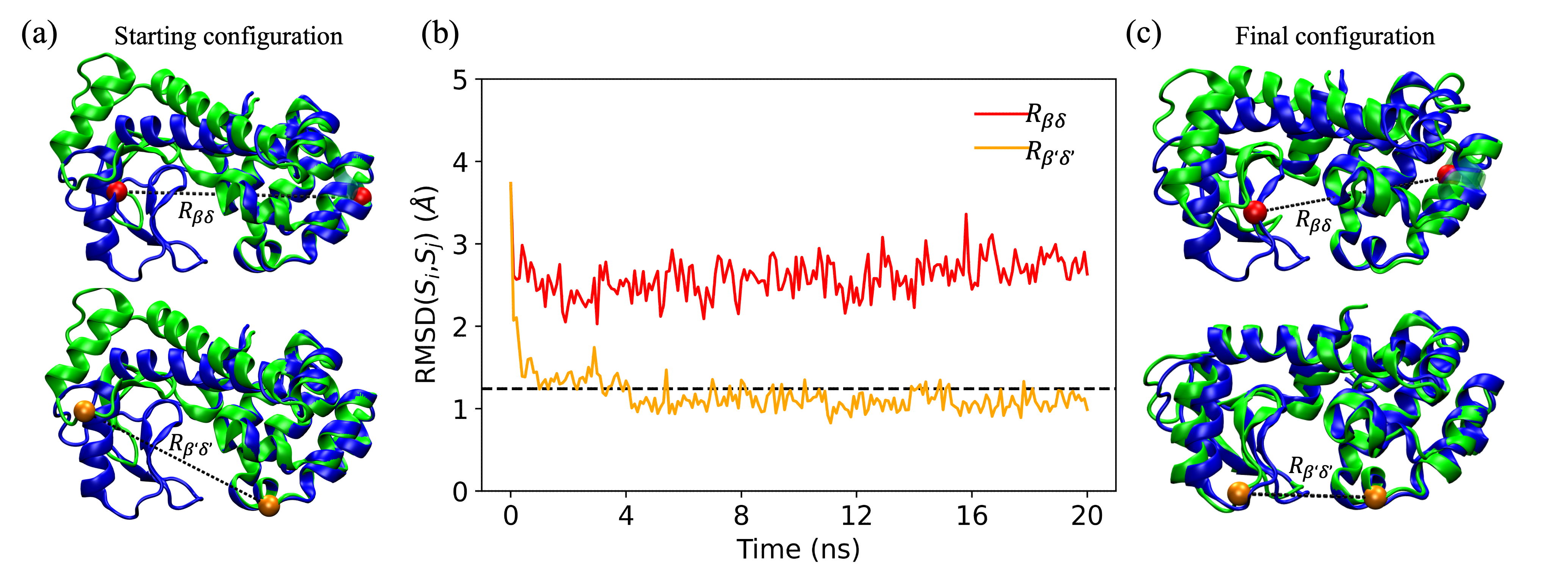}
\caption{(a) The top and bottom green images indicate the same starting structures for restrained MD simulations that enforce two different single restraints: $R_{\beta \delta} = R^0_{\beta \delta}$ (red) and $R_{\beta',\delta'}=R^0_{\beta',\delta'}$ (orange). The target structures, which are aligned with the starting structures such that the C$_{\alpha}$ RMSD values are minimized for residues $92$ to $162$, are shown in blue. (b) ${\rm RMSD}(S_i,S_j)$ as a function of time during restrained MD simulations for the restraints in (a), where the $S_i$ are structures sampled at each time during the restrained MD simulations and $S_j$ is the target structure. The black dashed line indicates ${\rm RMSD}_{r}$ of the target. (c) The final frames at $20$ ns from the restrained MD simulations with $R_{\beta \delta} = R^0_{\beta \delta}$ (top, red) and $R_{\beta',\delta'}=R^0_{\beta',\delta'}$ (bottom, orange). The structures from the final frames of the restrained MD simulations (green) are aligned with the target structure (blue) using the same alignment as in (a).}
\label{fig:2}
\end{figure}

Does the choice of the amino acid restraints have a significant impact on whether the initial protein structure can be moved to the target structure? As an example, we consider two possible choices for single amino acid restraints in restrained MD simulations of T4L. In Fig.~\ref{fig:2}, we plot the C$_{\alpha}$ ${\rm RMSD}(S_i,S_j)$ as a function of time, where $S_i$ are the structures sampled in the restrained MD simulations and $S_j$ is the target structure, for two possible choices for single amino acid restraints. Although the two restrained MD simulations start from the same initial structure (green) in Fig.~\ref{fig:2}a, the C$_{\alpha}$ ${\rm RMSD}(S_i,S_j)$ display different time dependence. The C$_{\alpha}$ RMSD from the MD simulations using the $(\beta',\delta')$ restraint rapidly decreases below the reference value, ${\rm RMSD}_r$, for the target structure. When we visualize the final frame from this restrained MD simulation in Fig.~\ref{fig:2}c, we find that the two subdomains of T4L are now closer together and there is strong alignment between the final frame (green) and target structure (blue). In contrast, the C$_{\alpha}$ RMSD does not decrease below RMSD$_r$ for the MD simulations with the $(\beta,\delta)$ restraint, even though the restraint is well-satisfied during the simulations. This example emphasizes the importance of the restraint selection method. Below, we describe the performance of four restraint selection methods in moving an initial structure toward a target structure.  We compare the performance of restraint selection methods that have information about the target structure and those that do not, and benchmark their performance to random restraint selection. In addition, we test the methods on moving four proteins from one {\it in vitro} structure to another, one protein from an {\it in vitro} structure to an in-cell structure, and the same protein from its in-cell structure to its {\it in vitro} structure.

\subsubsection*{Performance of restraint selection methods}

\begin{figure}[ht]
\centering
\includegraphics[width=1\linewidth]{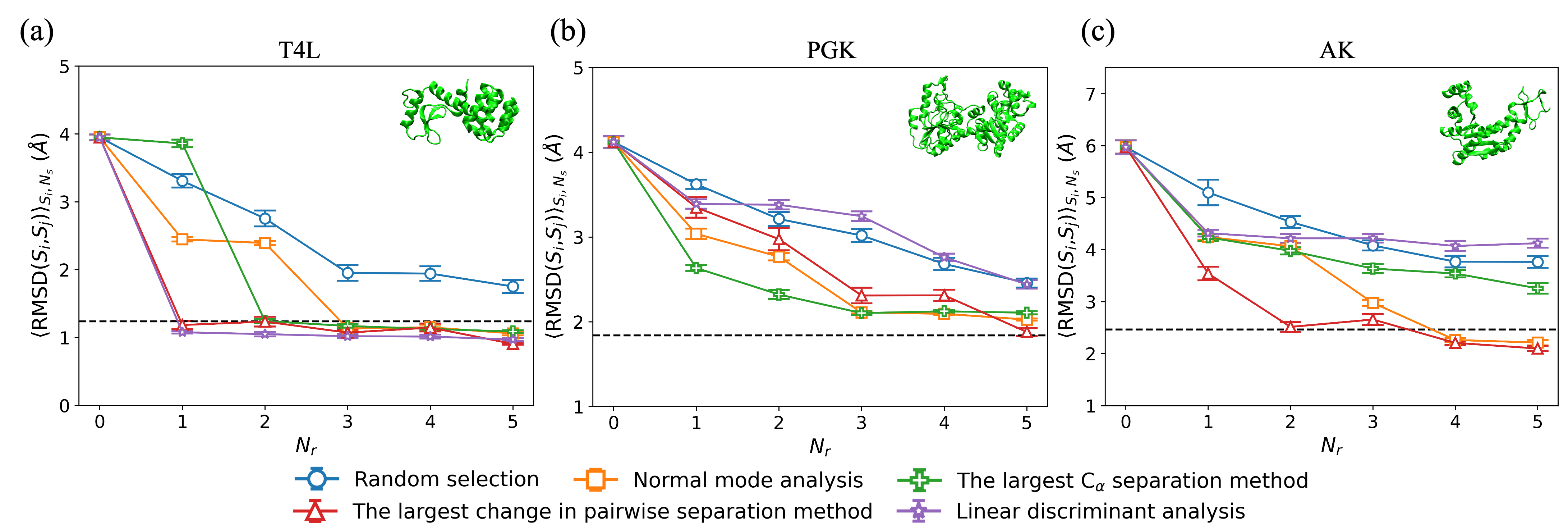}
\caption{The C$_\alpha$ RMSD, $\langle {\rm RMSD}(S_i,S_j) \rangle_{S_i,N_s}$, averaged over structures $S_i$ from the restrained MD simulations and number of samples $N_s$, where $S_j$ is the target structure, is plotted versus the number of restraints $N_{r}$ for (a) T4L, (b) PGK, and (c) AK. We show results for several restraint selection methods: random selection (blue circles), normal mode analysis (orange squares), the largest C$_{\alpha}$ separation method (green crosses), the largest change in pairwise separation method (red upward triangles), and linear discriminant analysis (purple stars). The horizontal black dashed line indicates ${\rm RMSD}_r$ for each target. Snapshots of the initial structures are shown for each protein in the upper right corner of each panel. The error bars give the standard error of $\langle {\rm RMSD}(S_i,S_j) \rangle_{S_i}$ from $N_s$ independent simulations.}
\label{fig:3}
\end{figure}

To quantify the performance of the four restraint selection methods in moving the initial structure to the target structure, in Fig.~\ref{fig:3}, we plot $\langle {\rm RMSD}(S_i,S_j) \rangle_{S_i,N_s}$ averaged over structures $S_i$ from the restrained MD simulations and number of samples $N_s$, where $S_j$ is the target structure, as a function of the number of restraints $N_r$ for several restraint selection methods and three proteins. In Fig.~\ref{fig:3}a, we show the results for T4L. When $N_r=0$ (i.e. unrestrained MD simulations), $\langle {\rm RMSD}(S_i,S_j) \rangle_{S_i,N_s} \sim 4$\AA~and the structures sampled in the MD simulations remain far from the target structure. When randomly selecting restraints, $\langle {\rm RMSD}(S_i,S_j) \rangle_{S_i,N_s}$ decreases slowly with increasing $N_r$ and does not reach ${\rm RMSD}_{r}$ even after five restraints have been included. In contrast, if we use structural information about the target to select the restraints (i.e. using linear discriminant analysis or identifying the largest change in pairwise C$_{\alpha}$ separations between the initial and target structures), $\langle {\rm RMSD}(S_i,S_j) \rangle_{S_i,N_s}$ rapidly decreases to ${\rm RMSD}_{r}$ after adding only one restraint. The normal mode analysis and largest C$_{\alpha}$ separation methods, which do not use information about the target, achieve intermediate results; $\langle {\rm RMSD}(S_i,S_j) \rangle_{S_i,N_s}$ decreases to ${\rm RMSD}_{r}$ within $N_r=2$-$3$ restraints. Thus, both of these methods (normal mode analysis and the largest C$_{\alpha}$ separation method) are more efficient than random selection for moving the initial structure to the target for T4L. 

To investigate the performance of restraint selection methods in moving an initial structure to the target structure for a wider range of proteins, we performed unrestrained and restrained MD simulations on two larger proteins, PGK (Fig.~\ref{fig:3}b) and AK (Fig.~\ref{fig:3}c), and find similar results for the variation of $\langle {\rm RMSD}(S_i,S_j) \rangle_{S_i,N_s}$ with $N_{r}$. (Note that for AK, we needed to add a background of intra-domain restraints to recapitulate the B-factor in the ``unrestrained" MD simulations as shown in Fig. S5 in Supporting Information.)  For example, for random restraint selection, $\langle {\rm RMSD}(S_i,S_j) \rangle_{S_i,N_s}$ slowly decreases with increasing $N_r$, and $\langle {\rm RMSD}(S_i,S_j) \rangle_{S_i,N_s} > {\rm RMSD}_r$ even for $N_r=5$. For PGK and AK, all restraint selection methods (except linear discriminant analysis) outperform random restraint selection. Normal mode analysis and the largest C$_{\alpha}$ separation method, which do not include information about the target, achieve lower values of $\langle {\rm RMSD}(S_i,S_j) \rangle_{S_i,N_s}$ than that for random selection, but their relative performance depends on the protein. For example, $\langle {\rm RMSD}(S_i,S_j) \rangle_{S_i,N_s}$ decreases faster with $N_r$ for the largest C$_{\alpha}$ separation method for PGK, whereas $\langle {\rm RMSD}(S_i,S_j) \rangle_{S_i,N_s}$ decreases faster for the normal mode analysis method for AK.  As expected, the method of identifying the largest change in pairwise C$_{\alpha}$ separations between the initial and target structures is the best performing method with $\langle {\rm RMSD}(S_i,S_j) \rangle_{S_i,N_s} \sim {\rm RMSD}_{r}$ after only a small number of restraints ($5$ for PGK and $2$ for AK). The performance of the restrained MD simulations based on the linear discriminant analysis method of selecting restraints is comparable to that for random restraint selection even though it incorporates structural information about the target. This result likely stems from the fact that a nonlinear method is needed to identify the collective variables in proteins.

\subsubsection*{Minimum fraction of restraints}

\begin{figure}[ht]
\centering
\includegraphics[width=0.8\linewidth]{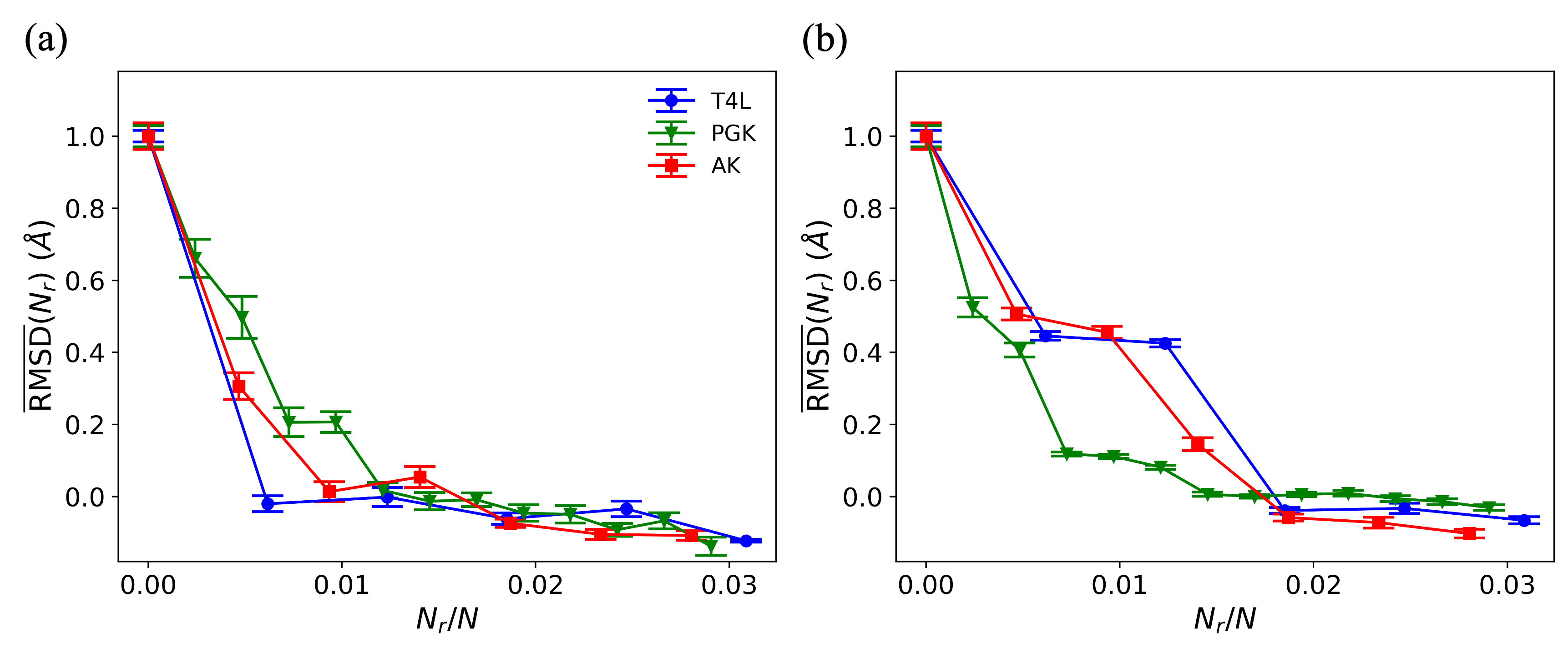}
\caption{The normalized C$_{\alpha}$ RMSD, $\overline{{\rm RMSD}}(N_r)$, is plotted versus $N_{r}/N$ for T4L (blue circles), PGK (green triangles), and AK (red squares) using (a) the largest change in pairwise C$_{\alpha}$ separation method and (b) the largest C$_{\alpha}$ separation method for selecting the restraints.}
\label{fig:4}
\end{figure}

These results demonstrate that the addition of a small number of pairwise distance restraints in restrained MD simulations can effectively move a protein from an initial structure to a target structure that is more than $4$ \AA~away. However, the minimal number of restraints required for $\langle {\rm RMSD}(S_i,S_j) \rangle_{S_i,N_s} \sim {\rm RMSD}_r$ depends on the protein.  What controls the minimum number of restraints? To address this question, we calculate the normalized C$_{\alpha}$ RMSD:
\begin{equation}
\overline{ {\rm RMSD}}(N_r) = \frac{ \langle {\rm RMSD}(S_i,S_j) \rangle_{S_i,N_s} (N_r) - {\rm RMSD}_{r}}{ \langle {\rm RMSD}(S_i,S_j) \rangle_{S_i,N_s}(0) - {\rm RMSD}_{r}},
\end{equation}
where $\langle {\rm RMSD}(S_i,S_j) \rangle_{S_i,N_s}(0)$ is $\langle {\rm RMSD}(S_i,S_j)\rangle_{S_i,N_s}$ from unrestrained MD simulations with $N_r=0$. Thus, $\overline{ {\rm RMSD}}(N_r)=1$ indicates that the protein samples the ensemble of initial structures and $\overline{ {\rm RMSD}}(N_r)=0$ indicates that the protein samples the target ensemble. In Fig.~\ref{fig:4}, we show that $\overline{ {\rm RMSD}}(N_r)$ collapses for the three proteins T4L, PGK and AK when plotted versus $N_r/N$ and using the optimal restraint selection method (i.e. the largest change in pairwise separation). In this case, only a small fraction of restraints, less than 1.5\% of the protein size, is required to move between the two {\it in vitro} protein structures.  For the normal mode analysis restraint selection method, which does not consider information about the target, the collapse of ${\overline {\rm RMSD}}$ with $N_r/N$ is not as complete, but ${\overline {\rm RMSD}} \sim 0$ for $N_r/N \gtrsim 0.02$ for all proteins. Thus, even in the absence of complete knowledge of the target structure, we can induce changes in protein structure from an initial {\it in vitro} structure to a target {\it in vitro} structure using only a small fraction of amino acid separation restraints.

\subsubsection*{Moving from initial {\it in vitro} structure to target in-cell structure for GB1}

\begin{figure}[ht]
\centering
\includegraphics[width=1\linewidth]{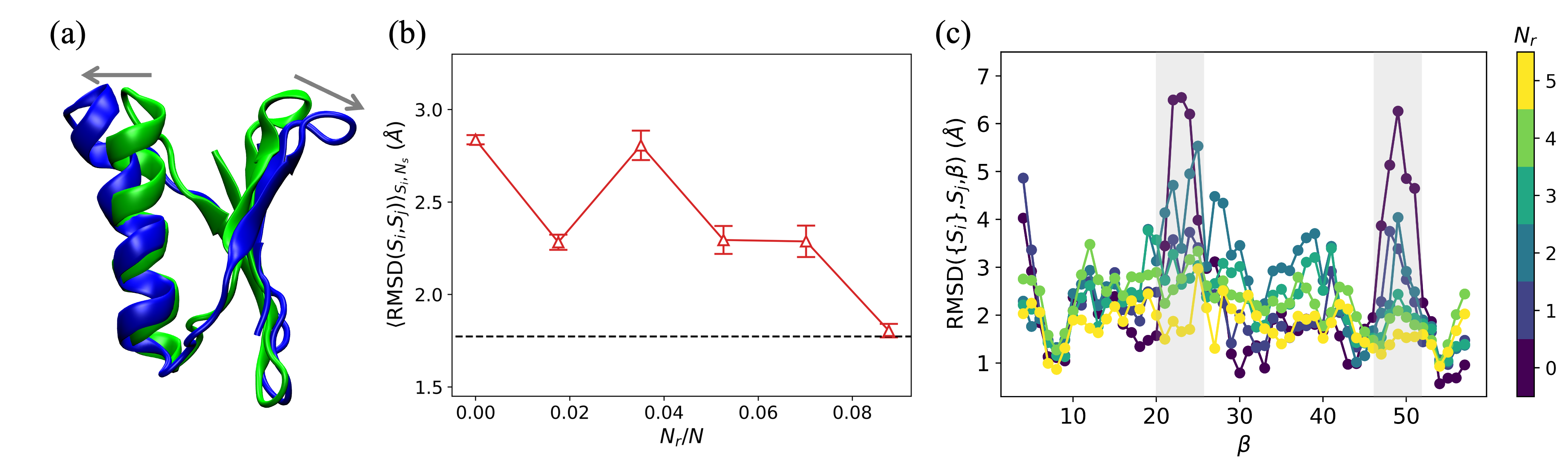}
\caption{(a) A ribbon diagram of the initial {\it in vitro} structure (green) of GB1 aligned with its in-cell target structure (blue). The gray arrows indicate the structural changes from the initial structure to the target structure in the two top loop regions. (b) $\langle {\rm RMSD}(S_i,S_j) \rangle_{S_i,N_s}(N_r)$ plotted as a function of $N_{r}/N$ for restrained MD simulations of GB1 using the largest change in pairwise C$_{\alpha}$ separation between the initial and target structures method (red upward triangles) for selecting restraints. The horizontal black dashed line indicates ${\rm RMSD}_{r}$ for the target. (c) ${\rm RMSD}(\{S_i\}, S_{j}, \beta)$ (Eq.~\ref{equation:3}) is plotted versus amino acid index $\beta$, where $\{S_i\}$ are structures sampled in the restrained MD simulations and $S_{j}$ is the target structure. The color from dark blue to yellow indicates $N_r$. The two top loop regions in (a) with $21 \leq \beta \leq 26$ and $47 \leq \beta \leq 52$ are highlighted in light gray.  }
\label{fig:5}
\end{figure}

To investigate whether the proposed methodology can also move an initial {\it in vitro} structure to a target {\it in vivo} structure, we also carried out unrestrained and restrained MD simulations for the B1 domain of protein G (GB1)~\cite{gerez2022protein}. We find that inter-bundle $\langle {\rm RMSD}(S_i,S_j) \rangle_{S_i,S_j} \sim 3.0$\AA~between the {\it in vitro} NMR models $S_i$ and the in-cell NMR models $S_j$ is rather small, mostly due to small changes in the two top loop regions shown in Fig.~\ref{fig:5}a. Despite this, the C$_{\alpha}$ RMSD between the {\it in vitro} and in-cell NMR bundles is larger than both of the intra-bundle fluctuations. Can we use similar restraint selection methods to those above to move from an {\it in vitro} structure to an in-cell structure? In contrast to the {\it in vitro} target structures for T4L, PGK, and AK, the in-cell target structure for GB1 is not a local minimum of the AMBER99SB-ILDN force field. (See Figs. S8a and b in the Supporting Information.) As a result, simulating in-cell protein structures is more challenging. In Fig.~\ref{fig:5}b, we implement the largest change in pairwise separation method for identifying restraints and calculate $\langle {\rm RMSD}(S_i,S_j)\rangle_{S_i,N_s}$ as a function of $N_r$. We find that the target in-cell structures are sampled for $N_r \ge 5$ in restrained MD simulations, which corresponds to $N_r/N \sim 8$\%. Unlike for the {\it in vitro} structures, where $\langle {\rm RMSD}(S_i,S_j)\rangle_{S_i,N_s}$ monotonically decreases as $N_{r}$ increases, $\langle {\rm RMSD}(S_i,S_j)\rangle_{S_i,N_s}$ exhibits non-monotonic behavior with $N_r$ for GB1. To investigate this effect, we calculate the C$_{\alpha}$ RMSD between the structures from the MD simulations and the target for each amino acid $\beta$ individually (Eq.~\ref{equation:3}). As shown in Fig.~\ref{fig:5}c, the deviations in the structures sampled in the unrestrained simulations and target structure occur predominantly in the top two loop regions in Fig.~\ref{fig:5}a with $21 \leq \beta \leq 26$ and $47 \leq \beta \leq 52$. The deviations between the initial and target structures can be decreased in the loop regions to $\lesssim 2$ \AA~for $N_r =5$. However, the structural deviations at amino acid positions adjacent to the loop regions, for example, $17 \leq \beta \leq 20$ and $27 \leq \beta \leq 31$, begin to increase as $N_r$ increases. These results suggest that the restraints are competing with the force field when attempting to move the protein to the in-cell structure. The non-monotonic behavior in  $\langle {\rm RMSD}(S_i,S_j) \rangle_{S_i,N_s}$ gives rise to a higher fraction of restraints that are necessary to move GB1 from the initial {\it in vitro} structure to the in-cell structure. 

\subsubsection*{Relation between minimum fraction of restraints and stability of target structure}

\begin{figure}[ht]
\centering
\includegraphics[width=0.8\linewidth]{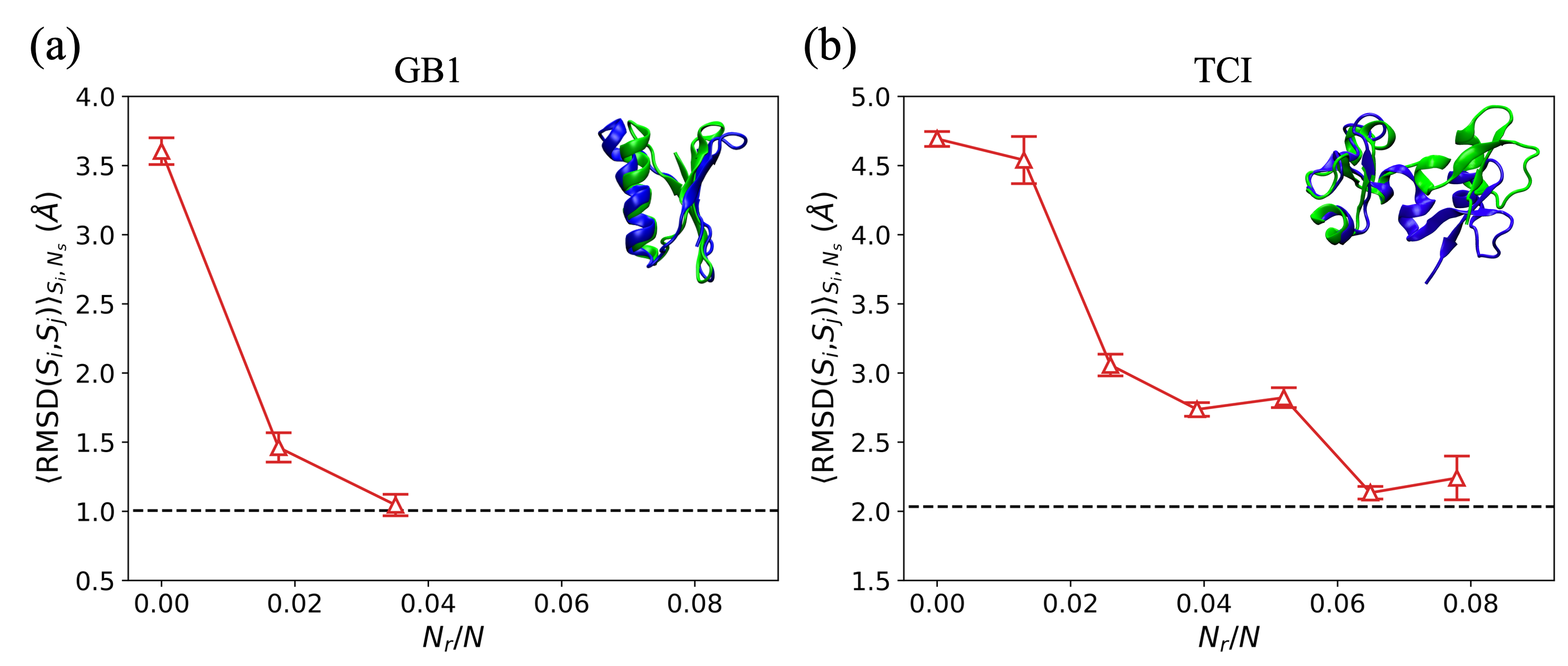}
\caption{$\langle {\rm RMSD}(S_i,S_j) \rangle_{S_i,N_s}$ plotted as a function of $N_{r}/N$ for restrained MD simulations of (a) GB1 and (b) TCI using the largest change in pairwise C$_{\alpha}$ separation between the initial and target structures method (red upward triangles) for selecting restraints. The horizontal black dashed lines indicate ${\rm RMSD}_{r}$ for the targets. Ribbon diagrams of the initial structures (blue) aligned with the target structures (green) for GB1 and TCI are shown in the upper right corners of (a) and (b), respectively. }
\label{fig:6}
\end{figure}

Several studies have suggested that GB1 is a single-domain protein, not a two-domain protein as for T4L, PGK, and AK~\cite{byeon2003protein}. It is possible that the larger fraction of restraints needed to induce a conformational change in GB1 (relative to the fraction needed for T4L, PGK, and AK) is related to the fact that it behaves like a single-domain protein. However, the variation in the minimum fraction of restraints can also be related to the stability of the target structure in the force field without restraints. To investigate this effect, we calculated the minimum number of restraints needed to move the ``unstable'' {\it in vivo} structure of GB1 to its ``metastable'' {\it in vitro} structure. We carried out restrained MD simulations of GB1 using restraints selected from the largest change in pairwise separation method. We find that only $\sim 3\%$ of the restraints are needed to move from the {\it in vivo} to the {\it in vitro} structure for GB1 (Fig.~\ref{fig:6}a), whereas $\sim 8\%$ of the restraints are needed to move from the {\it in vitro} to the {\it in vivo} structure. We also studied another two-domain protein Tick Carboxypeptidase Inhibitor (TCI) that undergoes a conformational change upon binding to its receptor. (The bound and unbound structures have PDBIDs 1ZLI and 2JTO, respectively.) The unbound NMR structure is unstable in the AMBER99SB-ILDN force field as shown in Fig. S7. We find that the fraction of restraints needed to move from the initial bound state to the target unbound state of TCI is significantly higher ($\sim 7\%$) than the fraction of restraints needed to move from the unstable {\it in vivo} state to the stable {\it in vitro} state of GB1, even though TCI is a two-domain protein. (See Fig.~\ref{fig:6}b.) Therefore, the characteristic fraction of restraints needed to induce conformational changes is primarily determined by the stability of the target state in the force field (without restraints).

\section{Discussion}
\label{discussion}

Numerous studies have emphasized that the in-cell environment can strongly influence protein structure and interactions~\cite{fulton1982crowded,ellis2001macromolecular,ebbinghaus2010protein,leeb2020diffusive,speer2021intracellular}. However, the complete atomic structure for proteins in the cellular environment has been obtained for only three proteins using NMR spectroscopy~\cite{sakakibara2009protein,tanaka2019high,gerez2022protein}. FRET has also been employed to determine a small number of separations between amino acids in several proteins {\it in vivo} ~\cite{ebbinghaus2010protein,wang2018cell,davis2018non}. In contrast, MD simulations, which have been calibrated to structures in the protein data bank, can provide the atomic coordinates of proteins in idealized, {\it in vitro} conditions. Properly calibrated force fields that would allow accurate all-atom descriptions of protein conformations {\it in vivo} currently do not exist. For example, GTT WW domain and three-helix bundle protein B (PB) fail to refold in all-atom cytoplasm computational models~\cite{rickard2020crowding,russell2023insilico}, since the interatomic sticking interactions in current protein force fields increase the stability of non-native states in all-atom cytoplasm models~\cite{samuel2023incell,rickard2019incell}. The restrained MD simulations of proteins {\it in vivo} (using residue separations measured in FRET experiments) can be carried out using current force fields, and thus we do not need to wait for improvements to the force fields in all-atom models of the cytoplasm.

While current all-atom MD simulations allow us to sample {\it in vitro} protein structures that match the x-ray crystal or NMR structures, we want to capture the all-atom conformational dynamics of {\it in vivo} protein structures. To do this, we can start with an initial {\it in vitro} protein structure, and add inter-residue distances measured by FRET as restraints in the MD simulations. However, an important question is how do we determine which amino acid pairs should be measured in the FRET experiments and then incorporated into the restrained MD simulations? In particular, what is the minimal number of restraints needed to achieve a given C$_{\alpha}$ RMSD to the target structure and what is the optimal method to select these restraints? To answer these questions, we first develop the methodology of restraint selection for proteins with multiple conformational states {\it in vitro}. To connect the {\it in vitro} results to those for {\it in vivo} conditions, we studied GB1, which is one of the few proteins whose structure has been solved {\it in vivo} using in-cell NMR spectroscopy. We selected the {\it in vitro} structure as the initial structure and the {\it in vivo} structure as the target structure. We employed four methods for selecting the restraints, varied the number $N_r$ and type of restraints that are incorporated into the restrained MD simulations, and compared the performance (i.e. C$_{\alpha}$ RMSD relative to the target structure) of the restraint selection methods to random selection. 

We found that for T4L, PGK and AK, which possess two distinct conformational states that have been solved {\it in vitro} and are both stable in the force field without restraints, it only takes a small fraction of restraints ($N_r/N<2\%$) to induce the conformational changes. The results emphasize that only a limited amount of information about pairwise distances between amino acids is needed to induce protein conformational changes from an initial structure to a target structure. The largest change in pairwise separation method, which uses the target structure information, gives the lowest values of C$_{\alpha}$ RMSD at each $N_r$. The normalized $\overline{ {\rm RMSD}}(N_r)$ for the largest change in pairwise separation method collapses for these three proteins when plotted versus $N_r/N$, and reaches zero at a small fraction of restraints ($N_r/N<1.5\%$). The linear discriminant analysis, which also uses the target structure information, exhibits inconsistent performance, sometimes comparable to and sometimes better than random selection. The normal mode analysis and the largest C$_\alpha$ separation method, which do not need target structure information, give lower values of C$_{\alpha}$ RMSD compared to random selection. The performance of these two methods varies slightly for different proteins. Specifically, for the normal mode analysis, $\overline{ {\rm RMSD}}(N_r)$ reaches zero when $N_r/N>2\%$, which is significantly lower than the fraction of restraints used in previous studies of FRET-assisted protein structural modeling (between $5\%$ and $12\%$)~\cite{dimura2020automated}. This result is significant since it shows that it is possible to determine the complete protein structure, using information from only $\sim 2\%$ of the C$_{\alpha}$ separations.

The studies described here provide proof of principle that FRET-assisted MD simulations can improve our understanding of the atomistic structure of proteins {\it in vivo} since only a small fraction of restraints are required to lock-in the target structures. We have shown that the characteristic fraction of restraints needed to induce conformational changes is determined by the stability of the target state in the force field without restraints. Since the {\it in vivo} structures are not stable minima of the current MD force fields, additional restraints are necessary to reach a given C$_{\alpha}$ RMSD and the dependence of the C$_{\alpha}$ RMSD on $N_r$ is non-monotonic. Therefore, an important future direction is to develop force fields for which the {\it in vivo} structures are potential energy minima. However, this goal requires a large number of high-quality, all-atom {\it in vivo} protein structures. In the absence of many {\it in vivo} protein structures, we can also develop in-cell mimetic systems whose crowding and surface sticking interactions have been calibrated to the {\it in vivo} studies~\cite{davis2018non,davis2020vitro}.

The linear discriminant analysis approach of using the direction that maximizes the differences in the initial and final state projections has been used successfully to identify the key pairwise separations that distinguish the initial and target states of small molecules~\cite{mendels2022collective}. However, while we find that linear discriminant analysis outperforms random selection only for T4L ($N=162$ amino acids), it performs worse than random selection for PGK ($N=413$) and AK ($N=214$). This result suggests that as the number of amino acids increases, the dimensionality of the conformation space increases to such an extent that the linear method cannot capture the key pairwise distances that describe the structural transition. Therefore, nonlinear methods, such as the nudged elastic band method, are needed to calculate the minimum energy pathway between the initial and final states and identify the most important pairwise distances along the pathway~\cite{jonsson1998nudged,ghoreishi2019fast}. 

The largest change in pairwise separation method that identifies a minimal set of amino acid pairs without redundant structural information gives the lowest values of C$_{\alpha}$ RMSD to the target structures. Note that one should not imagine a set of unique, optimal restraints, but an optimal pool of similar restraints that can induce structural changes (see Fig. S8 in Supporting Information). However, this method requires information about both the initial and target structures, and thus it cannot be used to {\it predict} an {\it in vivo} structural ensemble that has not yet been determined. Given that there are several NMR structures for proteins in cellular environments available, this selection method can be used in restrained MD simulations to verify that the simulations can correctly sample the conformational dynamics for these proteins {\it in vivo}. In contrast, if we aim to predict the {\it in vivo} structure of proteins, we have shown that normal mode analysis, which assumes that the normal modes of vibration of the initial structure provide information about how the protein transitions from the initial structure to the target structure, can be used to select the FRET-pair labeling positions that will enable the restrained MD simulations to sample the target {\it in vivo} structure. In this work, which considers large-scale domain motion in proteins, we showed that the normal mode analysis is a successful restraint selection method. However, it is not yet clear whether such motion is relevant for proteins {\it in vivo}. Thus, if the normal mode analysis is not efficient in inducing a change from the initial to the target {\it in vivo} structure, the largest C$_{\alpha}$ separation method, or a hybrid method that couples normal mode analysis and the largest C$_{\alpha}$ separation method, can also be implemented. (See Fig. S9 in Supporting Information.)

It is important to note that there are several experimental limitations for selecting amino acid pairs for FRET labeling. For example, the labeled dye molecules should not perturb the protein structure. Thus, in this study, we did not select amino acid pairs that occur within the core. FRET-labeling core amino acids would likely lead to changes in protein structure. In addition, optimal efficiency for FRET-labeling is achieved when the distance $d$ between donor and acceptor molecules is comparable to the characteristic F\"{o}rster distance $R_{0}$ for each dye pair. For most common FRET chromophores, $50$ \AA~$< R_0 < 60$~\AA, which is larger than $R_g$ for the proteins we considered in this work. Thus, we excluded amino acid pairs that are separated by $d < R_g$. 

As mentioned above, we showed in this study that the normal mode analysis selection method (and largest C$_{\alpha}$ separation method) can identify a small number of important C$_{\alpha}$-C$_{\alpha}$ distance restraints that can effectively move an initial {\it in vitro} structure to a target {\it in vivo} structure. However, since the FRET energy transfer efficiency reports on the ensemble of distances between dye molecules (not C$_{\alpha}$-C$_{\alpha}$ distances), quantitative FRET-assisted protein structural modeling requires the mapping between dye-dye distances and C$_{\alpha}$-C$_{\alpha}$ distances~\cite{klose2021resolving}. Therefore, in future studies, we will identify the mapping by incorporating atomic-scale modeling of the dye molecules into the restrained MD simulations. 

Finally, our current analysis of restraint selection methods focuses on monomeric, globular proteins. In future studies, we can expand the application of our restraint selection methods to intrinsically disordered proteins (see Fig. S10 in Supporting Information), membrane proteins, protein-protein and protein-nucleic acid complexes. Another interesting future direction is to develop restraint selection methods for triple restraints (co-restraints between 3 residues), using the inter-residue distances obtained from a 3-color FRET experiments~\cite{yoo2020fast}.\\

\textbf{Acknowledgements:} The authors acknowledge support from NIH Grant No. R35GM151146 (E. K. and C. D.) and NIH Training Grant Nos. T32GM145452 (Z. L., A. G., and C. O.) and T15LM007056-37 (J. S.), as well as the High Performance Computing facilities operated by Yale’s Center for Research Computing.\\

\textbf{Conflict of interest statement:} The authors declare no conflicts of interests.\\

\textbf{Data availability statement:} The code for implementing the five restraints selection methods can be found at \url{https://github.com/lzyttxs/restraint_selection_methods/}\\

\textbf{Supporting Information:} Additional material that supports the results and conclusions can be found online in the Supporting Information section at the end of this article.

\newpage


\end{document}